# Structural, electrical and energy storage properties of $BaO–Na_2O–Nb_2O_5–WO_3–P_2O_5$ glass–ceramics system

A. Ihyadn[a], A. Lahmar[b], D. Mezzane[a], L. Bih[c,d], A. Alimoussa[a], M. Amjoud[a], M. El Marssi[b], I.A. Luk'yanchuk[b].

[a]*Laboratory of Condensed Matter and Nanostructures, Cadi Ayyad University, Marrakesh, Morocco*

[b] *Condensed Matter Physics Laboratory, University of Picardie Jules Verne, Amiens, France*

[c]*Département Matériaux et Procédés, ENSAM Meknès, Université Moulay Ismail, Meknès, Maroc.*

[d] *Equipe Physico-Chimie la Matière Condensée (PCMC), Faculté des Sciences de Meknès, Maroc.*

ABSTRACT

Ferroelectric glass-ceramics are promising composite materials with dual properties, of high dielectric permittivity and high dielectric breakdown strength. In this context, new phosphate glass-ceramics $2BaO-0.5Na_2O–2.5[(1-x)Nb_2O_5–xWO_3]-1P_2O_5$ (x=0, 0.1, 0.2, 0.3 and 0.4) designated as B0, B1, B2, B3, B4 respectively, were prepared by the controlled crystallization technology and then crystallized at high temperature of 760 °C for 10 hours. Subsequently, the obtained glass-ceramics were studied by Differential Scanning Calorimetry (DSC), X-ray powder diffraction (XRD), Raman, impedance spectroscopy, and the P-E hysteresis loops. All of these techniques allowed the identification of $Ba_2Na_4W_2Nb_8O_{30}$ phase with tungsten bronze structure embedded in the glassy matrix. The density of glass-ceramics was found to increase with increasing tungsten content to reach a maximum at $x = 0.3$ and then decreases. Dielectric and conductivity parameters were found to be governed by the presence of oxygen vacancies. Room temperature energy storage property which depends on the composition X, was performed using *P-E* hysteresis loops of the glass-ceramics. The optimum discharge density was obtained for B3 with an energy efficiency of 73.77%.

## 1. Introduction

Glass-ceramic composite materials have been a subject of intensive interest during the last decades [1, 2, 3]. Obviously, the glass-ceramics are defined as composites containing at least one crystalline phase embedded in a glass matrix. Generally, these materials could be formed by the crystallization of the obtained glasses at certain temperatures [4]. The merit of such materials resides in the combination of the special properties of dielectric ceramics with the distinctive characteristics of glasses [4, 5]. It is worth noting that the solid state dielectric materials are propitious for electronic capacitors, however they exhibit low breakdown strength that imitates their application for energy storage. To overcome this shortcoming, the researchers were interested in the elaboration of dielectric glass-ceramics composites that could improve the breakdown strength by several orders of magnitude [5].

In this context, a large number of studies are focused on the niobate glass-ceramic system. Different types of ferroelectric glass-ceramic systems based on $SiO_2$ and/or $B_2O_3$ network such as $(BaO, Na_2O)–Nb_2O_5$ [6,7], $(SrO, BaO)–Nb_2O_5$ [8,9], $(Na_2O, SrO)–Nb_2O_5$ [10,11], $(K_2O, SrO, BaO)–Nb_2O_5$ [12,13,14] have been investigated widely. Furthermore, $(BaO, Na_2O)–Nb_2O_5–SiO_2$ system (BNNS) is a promising candidate in which $Ba_2NaNb_5O_{15}$ is the main crystalline phase, which combines a high dielectric constant (~240) and high Curie temperature of about 600°C [7,15]. Likewise, the phosphate glasses have the supremacy for the advantage technology due to their simple composition combined with the strong glass forming character, low glass transition temperature, and high thermal expansion coefficient [16, 17]. Nonetheless, a little number of researches were devoted to such glass–ceramics. Takashi et al. reported the crystallization and physical properties of the $(BaO, Na_2O)–Nb_2O_5–P_2O_5$ glass-ceramics [18, 19]. Likewise, in our previous work, Bih et al. studied $y(Ba_{2.15-x}Na_{0.7+x}Nb_{5-x}W_xO_{15})-(1-y)P_2O_5$ as new glass-ceramics for high energy storage density [20].
In the present study, a new glass-ceramic belong to the system $2BaO-0.5Na_2O–2.5[(1-x)Nb_2O_5–xWO_3]-1P_2O_5$ ($x$=0-0.4) (abbreviated as BNNWP) were prepared through controlling the crystallization technology. And the effects of the $WO_3$ substitution on thermal, structural, electrical properties and energy storage performances of BNNWP glass-ceramics.

## 2. Experimental procedures

The glasses $2BaO–0.5Na_2O–1P_2O_5–2.5[(1-x)Nb_2O_5- xWO_3]$ ($x$ = 0, 0.1, 0.2, 0.3, 0.4) were designated as B0, B1, B2, B3, B4 respectively, were synthesized using an alumina crucible by the conventional method of quenching. Firstly, analytical reagent grade powders of $Na_2CO_3$,

$BaCO_3$, $Nb_2O_5$, $WO_3$ and $(NH_4)H_2PO_4$ compounds were mixed for 1h in an agate mortar for homogenous mixing. The mixed powders were pre-annealed at 1000°C for 3 h to thoroughly remove ammonia and carbonate. The remaining mixture was pulverized again and then melted in the Al-crucible at 1350 °C for 40 min in the air. The melt was poured into a steel plate preheated and pressed with steel plate lid. Differential Scanning Calorimetry (DSC) (Model SDT-Q600) was used to predict the crystallization temperatures. The casting glass was then annealed for 1 h at Tg to relax the internal thermal stresses. The as-annealed glasses was cut and polished. The controlled crystallization of glasses sheets was carried out at the temperature of 760 °C for 10 h.. The X-ray diffraction (XRD) was conducted on Panalytical™ X-Pert Pro spectrometer having a Cu-Kα radiation (λ ~ 1.54 Å) to follow the phase evolution. The density (*d*) of the glass-ceramics was determined at room temperature by Densimeter Model H-300™ S, using the Archimede's method with distilled water as an immersion fluid. At least three samples of each glass composition were selected to determine the density. The molar volume *V* was calculated using the following formula (1):

$$V = \frac{M}{d} \qquad (1)$$

Where *M* is molecular mass of the glass-ceramics.

The structure of the glasses was studied at room temperature by Raman spectroscopy using confotec MR520 ™ with green excitation laser of 532 nm. The spectra were obtained in a backscattering geometry between 100 and 1300 $cm^{-1}$. The electrical and dielectric measurements were carried out on silver paint coated samples using an impedance analyzer (LCR meter hp 4284A 20Hz-1MHz). Finally the energy storage behavior of the BNNWP ceramics with different $WO_3$ contents was investigated using *P–E* hysteresis loops, which were measured by a ferroelectric tester (the TF Analyzer 3000) at ambient temperature and at the frequency of 1 kHz.

## 3. Results and discussion

### 3.1. Differential Scanning Calorimetry (DSC)

Figure 1 gives the DSC results of the B0, B1 and B3 glass samples obtained at a heating rate of 10°C $min^{-1}$. It is clearly shown from the figure that the glass transition temperature (*Tg*) is slightly shifted to low temperature (from 683°C to 649 °C) with increasing tungsten content. Such decrease of *Tg* indicates that the replacement of $Nb_2O_5$ by $WO_3$ induces the depolymerisation of the network. The reason for a decrease in Tg is the differences in the energy of Nb–O and W–O bonds in the glass structure. The replacement of stronger Nb-O bonds ($E_{Nb-O}$ =771.5 kJ/mol [21]) by weaker W-O bonds ($E_{W-O}$ =672.0 kJ/mol [22]) contributes to the observed steep decrease of Tg.

Furthermore, DSC plots present two exothermic peaks, *Tp1* observed in the temperature range of 730 °C- 755 °C and *Tp2* situated in the interval of 798 °C- 825 °C. The presence of such two anomalies is a signature of the formation of two crystalline phases from the glass matrix. Moreover, B1 glass present a supplementary third peak *Tp3 at 780 °C*, that could be ascribes to a third crystalline phase.

According to the DSC analysis, the glass-ceramics were heat treated at 760°C to obtain crystallized phases with relatively high dielectric constant.

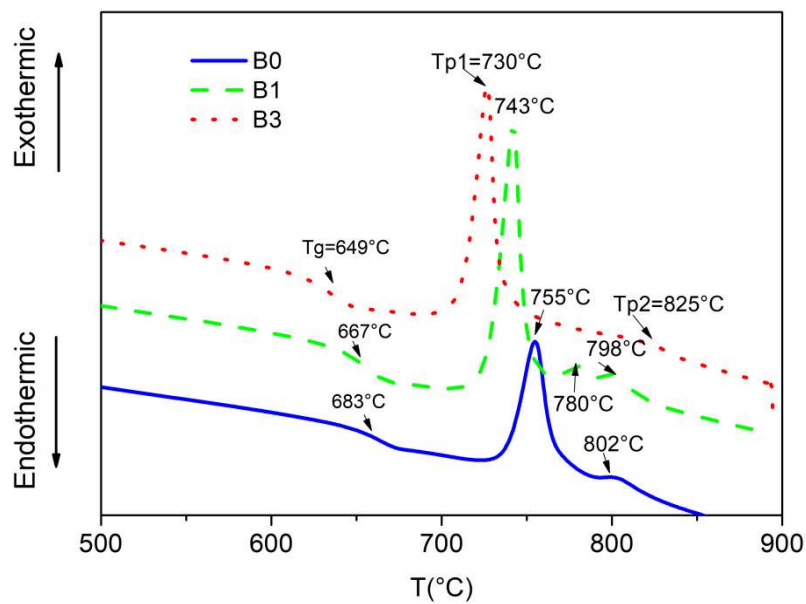

**Figure1.** Differential Scanning Calorimetry (DSC) curves of the BNNWP glasses.

*3.2. Density and molar volume determination*

The variation of the density and molar volume as a function of the composition are given in figure 2. The plots show the existence of an optimum around $x$ = 0.3 where the density was found to increase from 3.66g/cm$^3$ (for $x$ = 0) to 4.52 g/cm$^3$ ($x$ = 0.3) then decreases to 3.67g/cm$^3$ for $x$=0.4. This increase could be associated with the replacement of the lighter niobium ions (92.92 g/mol) by the heavier tungsten ions (183.85 g/mol) in the glass matrix. Besides, the molar volume presents an opposite trend for the density. The decrease of the molar volume could be interpreted in terms of decreasing of the mean ionic radius related to the substitution of $Nb^{5+}$ by $W^{6+}$, since $r(Nb^{5+})$ =0.64 Å and $r(W^{6+})$ =0.6 Å.

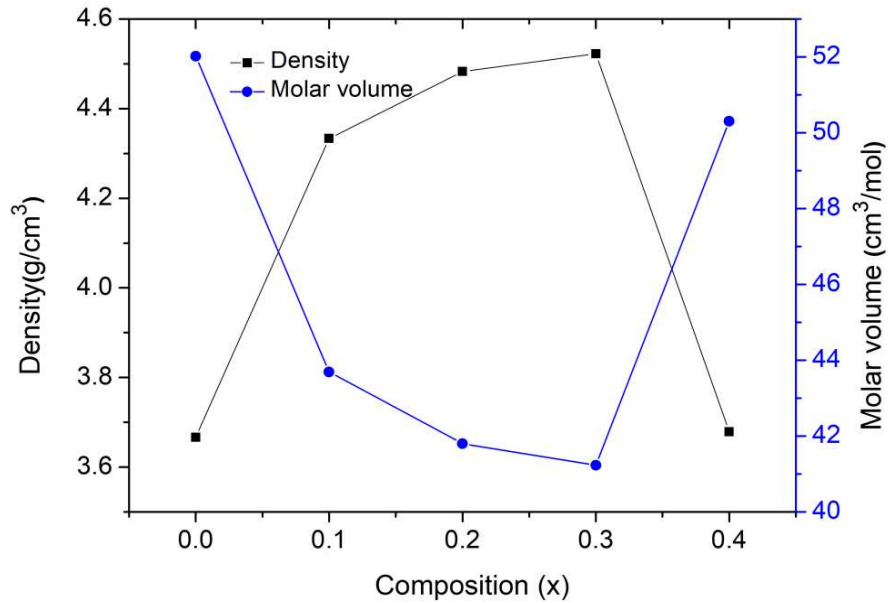

**Figure 2.** Variation of the density and the molar volume of glass-ceramics as a function of the tungsten content (x).

On the other hand, the number of oxygen anions increases gradually in the glass network with the increasing $WO_3$ content (B4). This means that the addition of $WO_3$ can promote a relatively open structure causing an increase in the molar volume. This observed behavior suggests that the molecular volume is influenced by the number of $O^{2-}$ ions per unit of volume.. The increase of molar volume in tungsten-glass-ceramics means that the excess of oxygen induces the formation of additional non-bridging oxygen. The same results were observed elsewhere [23]. Moreover, this decrease is explained by the XRD results as described below.

*3.3. Room temperature X- ray diffraction analysis (XRD)*

The phase evolution of BNNWP glass-ceramics as a function of $WO_3$ content was analyzed by XRD technique. As shown in figure 3, XRD diffractograms reveal the formation of crystalline phases after the heat treatment at 760°C. For B0 sample, the obtained reflections could be indexed mostly in orthorhombic tungsten bronze (OTB) isotype to $Ba_2NaNb_5O_{15}$ phase. The presence of a secondary phase, isotype to the orthorhombic $NaNbO_3$ perovskite, is also detectable . For B1, B2, and B3 samples, $Ba_2Na_4W_2Nb_8O_{30}$ and $NaNbO_3$ phases with tetragonal tungsten bronze (TTB) monoclinic structures respectively were identified. However, for the B4 sample, two orthorhombic TTB phases were spotted: $Ba_2NaNb_5O_{15}$ and $Ba_2Na_4W_2Nb_8O_{30,}$ in addition to the orthorhombic perovskite phase $NaNbO_3$.

The $Ba_2Na_4W_2Nb_8O_{30}$ phase was firstly reported by Ikeda et al [24] with TTB-type structure (a= 12.39 Å; c= 3.93 Å) and with a transition temperature around 365 °C, it could be derived directly from $Ba_2NaNb_5O_{15}$ by the substitution of $Nb^{5+}$ (0.64 Å, CN = 6) by $W^{6+}$ (0.60 Å, CN = 6) in the respect of electrical neutrality as following [25]:

$$Nb^{5+} + Ba^{2+} = W^{6+} + Na^+$$

It is worth noting that the XRD peak around 28° shifts toward higher angles as x increases up to 0.3 and then shifts to lower angles for x=0.4 (figure 3). This indicate that the lattice volume fraction decrease for B4, and increase for other samples, which is agreement with the variation of molar volume observed in figure 2. The observed high shift in XRD peak at 28° is due to the fact that the size of doping element $W^{6+}$ ($r(W^{6+})$ =0.6Å) is smaller than its substituting element $Nb^{5+}$ ($r(Nb^{5+})$ =0.64 Å).

Supplementary structural information, not always accessible by XRD, could be obtained using Raman spectroscopy.

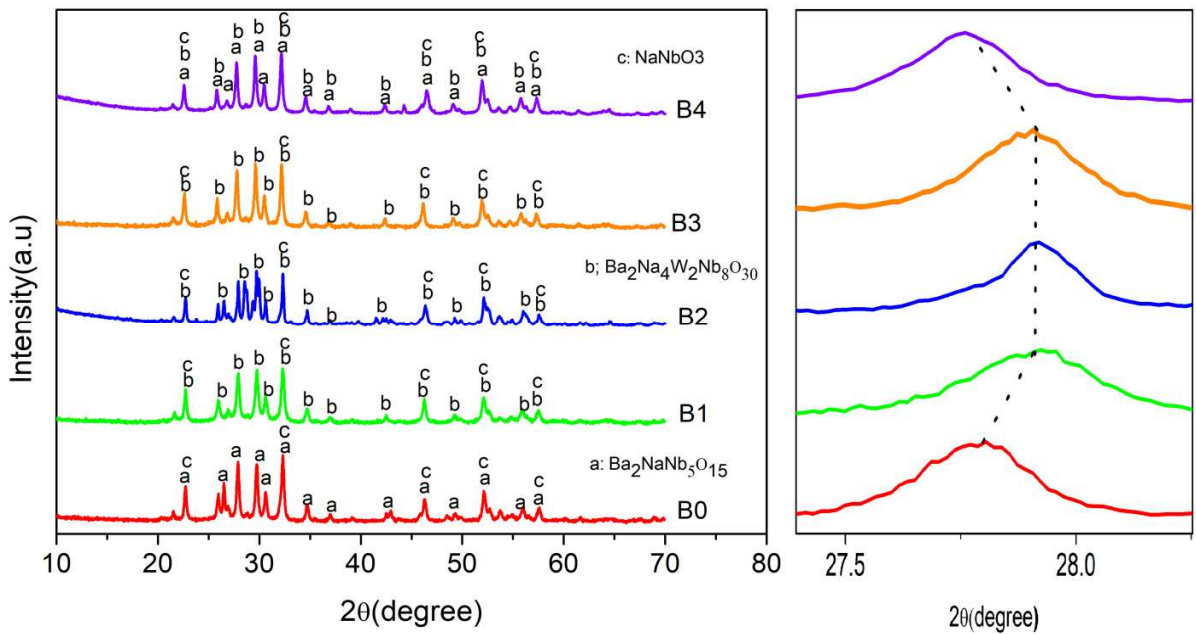

**Figure 3**. XRD patterns of the glass-ceramics specimens (in the left). In the right, the diffractogram shows shift in the position of the reflection around 28°.

*3.4. Raman spectroscopy study*

In the present work, Raman spectroscopy was conducted to characterize the molecular composition and the structure of the obtaining composite glass- ceramics [26]. The evolution of Raman spectra of prepared glass-ceramics is presented in figure 4.

The strong bands at around 630 cm$^{-1}$ are attributed to $\upsilon s$(M-O) vibrations in $NbO_6$ (M = Nb, W) and/or to $\upsilon s$(P-O-P) vibrations[27, 28, 29]. The three bands at 123; 250; 350 cm$^{-1}$, located in the frequency range of 100–400 cm$^{-1}$, are associated with the deformation modes of $MO_6$ units and δ(O-P-O) in PO$_4$ tetrahedra [23,29,30]. The bands at around 720 cm$^{-1}$ are due to bridging oxygen P–O–P stretching modes [31, 32, 33]. The band at 915 cm$^{-1}$ can be ascribed to the distorted [NbO$_6$] octahedral having at least a short Nb–O bond pointing towards a modifying ion [23,34,35]. The summary of the data on the positions of various bands of Raman spectra assignments is presented in Table 1. With the increase of tungsten content, the intensity of the band at 635 cm$^{-1}$ decreases. This could be explained by the decrease of symmetrical octahedral [NbO$_6$] sites.

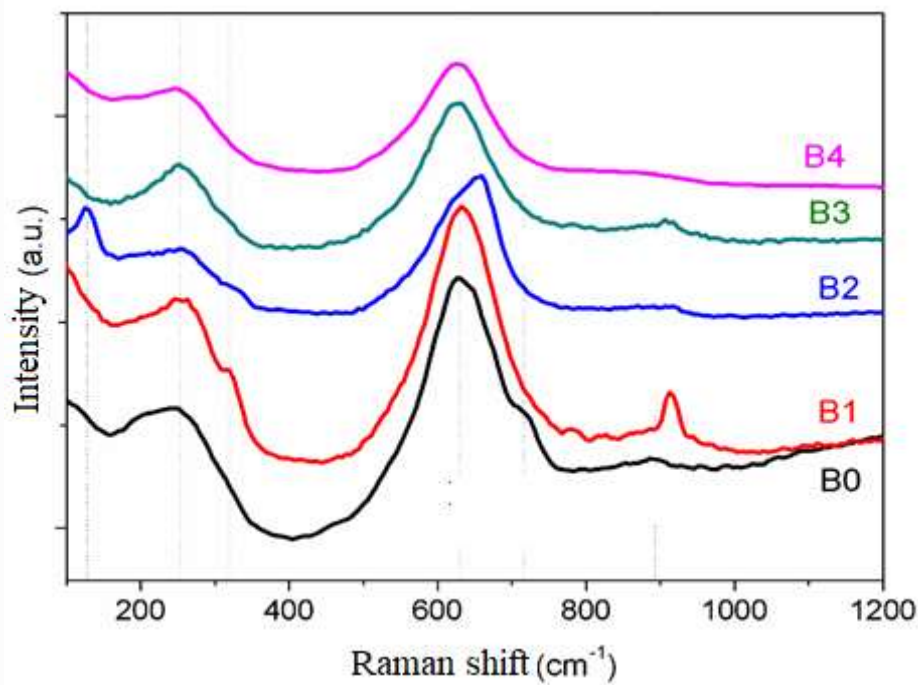

**Figure 4.** Raman spectra of the investigated samples in the frequency range 100-1200 cm$^{-1}$.

**Table 1:** Raman band positions of the glass-ceramics and their assignments.

|    | $\upsilon$(NbO$_6$)-δ(PO$_4$) | $\upsilon$(NbO$_6$) | $\upsilon$(NbO$_6$) | $\upsilon$(NbO$_6$) | $\upsilon_s$(P-O-P) | (Nb-O)/(W-O) |
|----|---|---|---|---|---|---|
| B0 | - | 234 | - | 630 | 723 | 893 |
| B1 | - | 252 | 323 | 632 | 716 | 914 |
| B2 | 131 | 247 | 325 | 658 | 714 | 913 |
| B3 | - | 251 | 321 | 630 | 713 | 910 |
| B4 | - | 248 | - | 629 | 711 | 875 |

The two highest frequency bands near 282 to 650 cm$^{-1}$ frequencies are apparently related to the stretching and bending vibrations of the NbO$_6$ octahedrons respectively. This is similar to the 630-650 cm$^{-1}$ and 260-280 cm$^{-1}$ bands, which are observed for the Ba$_{2,15-x}$Na$_{0,7+x}$Nb$_{5-x}$W$_x$O$_{15}$ (0≤x≤1) solid solution [36, 18]. The substitution seems to lead to the precipitation of a new phase (TTB): Ba$_2$Na$_4$W$_2$Nb$_8$O$_{30}$.

The appearance of bands near 916 cm$^{-1}$ can be ascribed to the incorporation of tungsten ions into the phosphate network and the formation of mixed structural units of the type P-O-W, W-O-W linkages [23]. The addition of $WO_3$ promotes an increase in the number of W–O bonds, which present at higher polarizability when compared to the P–O bonds, masking the spectrum associated with the entities of phosphate structural units. Due to the fact that the Nb–O and W–O bonds possess both similar atomic properties and high polarizability, the bands of the vibrations of Nb–O and W–O overlap and appear in the same frequency positions of the spectra [23].

*3.5. Dielectric and Electrical conductivity investigations*

The temperature dependence of the dielectric constant ($\varepsilon'$) for the different glass samples at 10 kHz is shown in figure 5. It is observed that the $\varepsilon'$ increases with temperature for B0, B2 and B3 samples without the observation of any anomaly. It seems that the high conductivity hides the observation of eventual anomalies. Moreover, it is worth mentioning that $Ba_2NaNb_5O_{15}$ phase is a particular composition of a solid solution of the tetragonal tungsten bronze structure (TTB) involving a wide domain of existence limited by $Ba_{1.90}Na_{1.2}Nb_5O_{15}$ and $Ba_{2.27}Na_{0.46}Nb_5O_{15}$ compositions [37]. Theses phases are non-stoichiometric and vacancies are part of the structure. For example, in the $Ba_{2.27}Na_{0.46}Nb_5O_{15}$ phase, the pentagonal and square sites are partially occupied, allowing to appear a proportion of 9% of vacancies. We believe that the increase of $\varepsilon'$ with the temperature is governed by such vacancies. The contribution of space charge polarization could be expected also, since other contributions for dielectric constant like electronic and orientation polarizations do not increase with temperature [38].

However, B1 and B4 glass-ceramics show the presence of transition peaks. For B1 sample, a dielectric anomaly is detected around 260°C. Recall that this sample is composed by $Ba_2NaNb_5O_{15}$ and $Ba_2Na_4W_2Nb_8O_{30}$ tetragonal tungsten bronze structures.

Both of these phases exhibit phase transition from orthorhombic to tetragonal symmetry around such temperature [37]. Taking a close look to the dielectric plot of the sample B4, one can depict an abnormal broad phase transition centered at 190°C. As we reported before on the structural characterization part, $Ba_2NaNb_5O_{15}$, $Ba_2Na_4W_2Nb_8O_{30}$, and $NaNbO_3$ are the main phases constitute B4. Furthermore, $NaNbO_3$ exhibits a phase transition around 142°C [39]. We can surmise then, that the large dielectric anomaly is caused by the superposition of two transitions, that one at 142°C is associated to $NaNbO_3$, and the other at 260°C is related to TTB phases

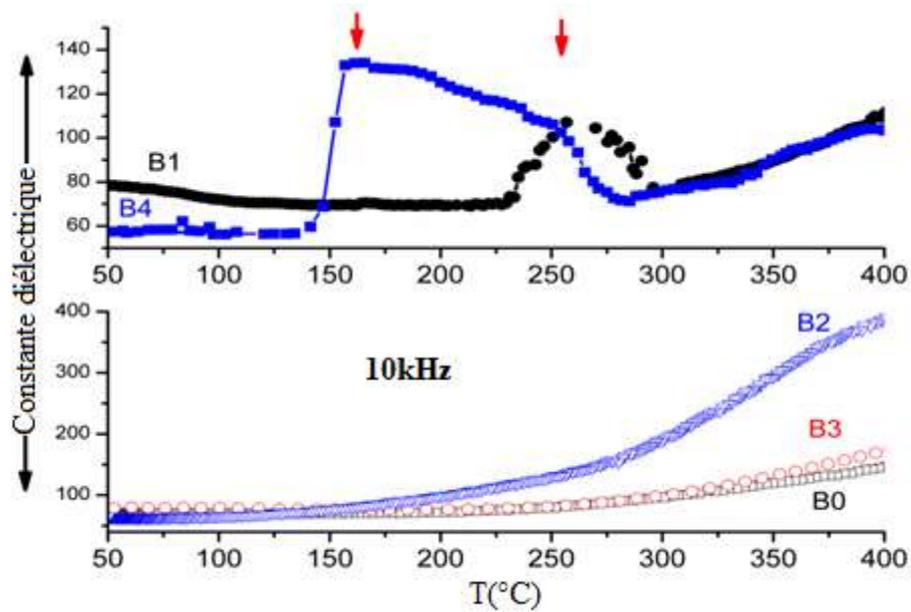

**Figure 5.** Variation of (a) the dielectric constant as function of temperature for the different samples.

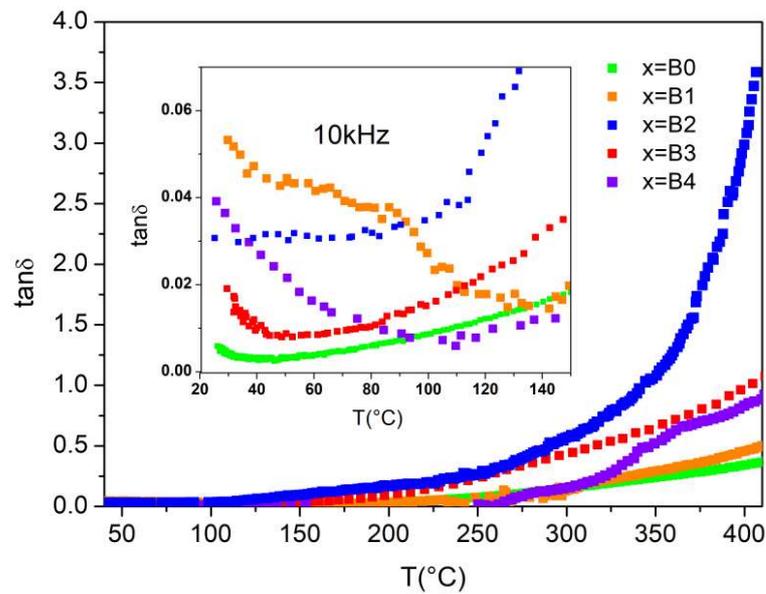

**Figure 6.** Variation of the tanδ as function of temperature for the different samples of glass-ceramics

The temperature dependence of the dielectric loss for the different glass samples at 10 kHz is shown in figure 6. It is noted that the dielectric loss, tan$\delta$, presented the same behavior of constant dielectric at high temperature. At room temperature, it is observed that the dielectric loss has a minimum value of 0.0055 corresponding to B0 sample which increased to a maximum (0, 0485) then decreased to 0, 0032 and 0,019 for B2 and B3 respectively. After it increased again to 0,039 for B4 sample.

It observed from Raman spectra (see Fig.4) that the intensity of the band of B1 centered at 914 cm$^{-1}$ (due to deformed WO$_4$ polyhedra) decreased with the increase of WO$_3$, which means a transformation of WO$_4$ to WO$_6$ groups [38]. This indicates that some of W ions are present in the W$^{6+}$ state. Then, the electron hopping may not possible between W$^{6+}$ and W$^{6+}$ ions and these ions makes W$^{6+}$–O–W$^{6+}$ chain inactive and it may not contribute to the conduction process , which can be linked to the decrease of dielectric loss of B2 and B3. For B4 the conductivity increases with the increasing of non-bridging oxygen (NBO) because the formation of NBOs would make the glass structure more open. This result is observed from the variation of the density and the molar volume in figure 2.

The *ac* conductivity ($\sigma_{ac}$) was calculated at different temperatures using the equation (2) [40,27]:

$$\sigma_{ac} = \omega \varepsilon_0 \varepsilon' \tan \delta \quad (2)$$

where ω=2πf, f represents the frequency, and ε$_0$ is the vacuum dielectric constant.

Figure 7 (a) shows as an example of the frequency dependent conductivity, ($\sigma_{ac}$) of B3 sample at different temperature. The other samples present the similar conductivity behavior (not shown here). The data obtained at different temperatures show two distinct behaviors within the measured frequency range. The first one is at low frequency with an almost constant evolution (plateau behavior) and the second present high frequency dispersion. The plateau region corresponding to *dc* conductivity, is found to extend to higher frequencies when temperature increases [41]. Noting that the existence of a non linear behavior is already pointed out as a signature of the presence of mixed conductivity, namely ionic and electronic ones [42,27]. The frequency at which this dispersion takes place ($\omega_h$) is temperature dependent and the reciprocal temperature dependence of $\omega_h$ follows the Arrhenius relation:

$$\omega = \omega_0 \exp\left(-\frac{Ea}{TK_B}\right) \quad (3)$$

with $\omega_0$ is a pre-exponential factor; $K_B$ is the Boltzmann constant, and $E_a$ represents the activation energy. The observed frequency independent conductivity (figure 7 (a)) could be arise from random diffusion of the charge carriers via activated hopping process, where only successful diffusion contributes to the *dc* conductivity. In contrast, the strong frequency dispersion of the conductivity is due to inhomogeneous in the glass-ceramics in a microscopic scale, where the distribution of the relaxation processes manifests through the distribution of energy barriers [43].

A typical $\ln\omega_h$ versus (1/T) plot is linear as shown in figure 7 (b) to determine the activation energy of the hopping process. It is worthy to note that the conductivity scale behavior has been widely studied using jump frequency as reduction parameter for the frequency axis, where $\omega_h$ takes into account the permittivity change and correlation effects [44]. The values of the activation energy are founded ranging between 0.19 to 1.05 eV, in the order of glassy materials where a small hoping polaron is mostly reported as the predominated conduction mechanism [45].

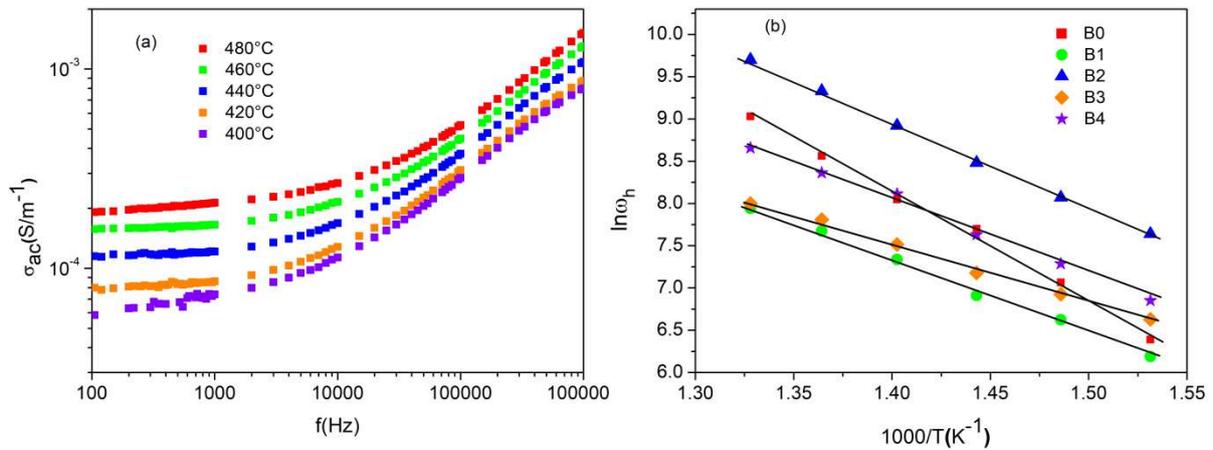

**Figure 7.** (a) Frequency dependent $a_c$ conductivity of B3 glass-ceramic at different temperatures, (b) temperature dependence of $\omega_h$ for BNNWP samples.

In the figure 8 (a) and (b), the *dc* and *ac* conductivity (logarithmic scale) behavior as a function of the 1000/*T* for all studied samples are plotted, respectively. It can be seen from the plots, both of *ac* and *dc* conductivity seems to follow a linear evolution with the rise of the temperature. This increase can be correlated with the thermal activated process. The obtained plots are found to be linear following the Arrhenius equation given by [46]:

$$\sigma = \sigma_0 \exp\left(-\frac{E_a}{K_B T}\right) \quad (4)$$

where $\sigma_0$ is a pre-exponential factor, $K_B$ is the Boltzmann constant and $E_{a\ ac}$ is the activation energy. All activation energies were calculated from the slope of the fitted line for the BNNWP glass-ceramics. The founded values are gathered in Table 2.

Except for B4 sample, the conductivity decrease with the increase of the activation energy. Such behavior is in good agreement with the Mott general formula [47] and consistent with literature reported work [48]. All these glass-ceramics possess electrical conductivity *dc* ranging from 9.98x $10^{-6}$ to 6.03x $10^{-4}$ (S.m$^{-1}$) at temperatures from 673 K and are in agreement with the reported *dc* conductivities in other glass-ceramics systems [27, 49].

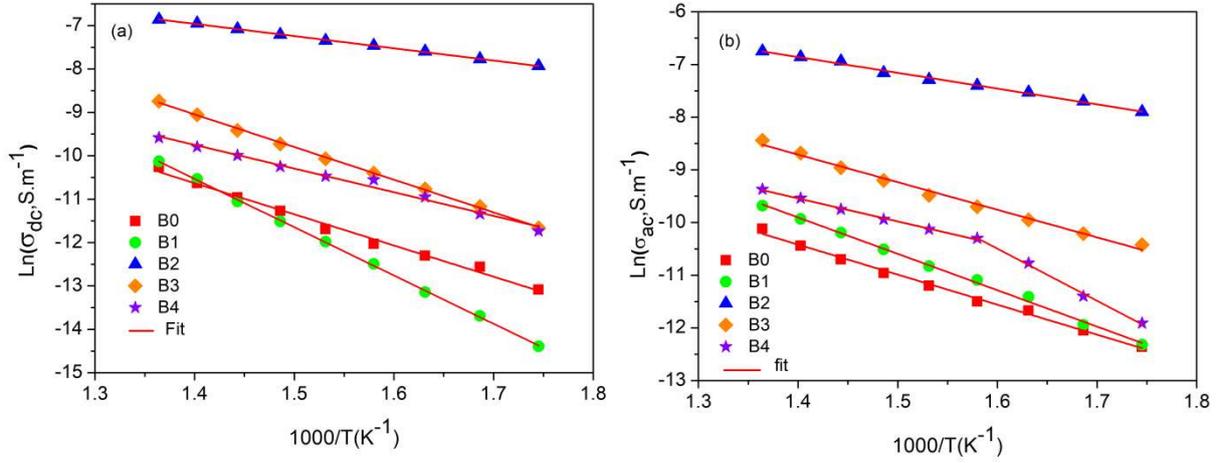

**Figure 8.** Arrhenius plot of BNNWP glass-ceramics at 10 KHz: (a) log ($\sigma_{dc}$) vs. 1000/T, (b) log ($\sigma_{ac}$) vs. 1000/T at 10 kHz.

Regarding to the influence of WO$_3$ content, it is observed that the conductivity decreased to a minimum with a small amount addition of tungsten oxide (*x*= 0.1). As we have mentioned above, the formed Ba$_2$NaNb$_5$O$_{15}$ is a lacunar structure, thus the incorporation of tungsten could be acted as donor doping ($W_{Nb}^{\cdot} + 1e^{-}$) leading to the oxygen vacancy annihilation, similarly to what is known for bulk materials [50].

**Table 2:** Dc conductivity at 673 K ($\sigma_{dc}$) and dc activation energy (E$_{a\,dc}$) calculated by the Arrhenius equation, ac conductivity ($\sigma_{ac}$) (at 673 K and 10 kHz), ac activation energy (E$_{a\,ac}$).

|    | $\sigma_{dc}$ (S/m) (673K) | $E_{a\,dc}$(eV) | $\sigma_{ac}$ (S/m) (673K, 10kHz) | $E_{a\,ac}$ (eV)(10kHz) | $E_{a(\omega h)}$(eV) |
|----|---|---|---|---|---|
| B0 | 1.27x 10$^{-5}$ | 0.72 | 1.89x 10$^{-5}$ | 0.48 | 1.05 |
| B1 | 9.98 x 10$^{-6}$ | 0.96 | 2.83x 10$^{-5}$ | 0.45 | 0.76 |
| B2 | 6.03 x 10$^{-4}$ | 0.23 | 7.7 x 10$^{-4}$ | 0.22 | 0.19 |
| B3 | 5.96x 10$^{-5}$ | 0.65 | 9.90x 10$^{-5}$ | 0.38 | 0.6 |
| B4 | 3.54 x 10$^{-5}$ | 0.47 | 4.79x 10$^{-5}$ | 0.41/0.84 | 0.76 |

An increase of WO$_3$ to *x*=0.2, the conductivity increase and reach a maximum value in both *dc* and *ac,* then decreased for *x* ≥ 0.3. This behavior could be linked to a change in the conduction mechanism from ionic to electronic one. Noting that the W–O–W bonds can be present either as W$^{6+}$–O–W$^{6+}$ or W$^{6+}$–O–W$^{5+}$ [27,43]. The creation of a large number of W$^{5+}$-O-W$^{6+}$ bonds can promote the electronic conductivity [51], which can explain the values obtained for *x*=0.2. Further increase of WO$_3$ concentration, may lead to the formation of a non-bridging oxygen (NBO) acting as charge compensation centers as reported by Upender et al [52]. The presence of such NBOs impedes the motion of W, inducing thus the decrease of the conductivity.

Noting that the case of sample B4 is very intriguing, as it's the only sample that presented a deviation from linearity in the *ac* conductivity. This behavior suggests the existence of a supplementary mechanism of electrical conductivity. According the activation energy values determined from both slops (see Table 2), we note that the first one of 0.41 eV is comparable to the rest of samples and could linked to the similar mechanism. However, the second energy activation is very big and it is around 0.84 eV. Such value is characteristic for ionized oxygen vacancy in the bulk like materials [53]. Noting that the oxygen ions hopping between sites could be plausible if we take in consideration the presence of different effective valence charges on the O-sites ($W^{6+}$, $W^{5+}$, $Nb^{4+}$, $Nb^{5+}$). Furthermore, the formation of NBOs (because the increase of $WO_3$ in glassy matrix) would make the glass structure more open with the reduction of the jump distance of the mobile ions [52].

*3.6. Energy storage determination*

Generally, the energy storage properties of the dielectric materials are characterized by recoverable energy density ($W_{rec}$), energy loss density ($W_{loss}$), and energy efficiency ($\eta$), which can be calculated through the hysteresis loop. Among them, $W_{rec}$ is equal to integral of the area between the polarization axis and the discharge curve in *P–E* hysteresis loops while the area of the *P–E* loop represents is $W_{loss}$. The three parameters are computed by using following equations:

$$W_{rec} = \int_{Pr}^{Pmax} EdP \qquad (5)$$

$$W_{loss} = \int_{0}^{Pmax} EdP - W_{rec} \qquad (6)$$

$$\eta = \frac{W_{rec}}{W_{rec}+W_{loss}} \times 100 \qquad (7)$$

The ambient temperature polarization–electric field (*P–E*) hysteresis loops for the BNNWP glass-ceramics measured at 90 kV/cm and 1 kHz are shown in figure 9 (a). While figure 9 (b) presents the schematic diagram for calculation of energy storage density using the hysteresis loop, in which the red area is the discharge energy density and the blue region designates to the energy loss density.

The recoverable energy ($W_{rec}$), the energy loss density ($W_{loss}$) and the efficiency ($\eta$) obtained from *P-E* loops are listed in table 3.

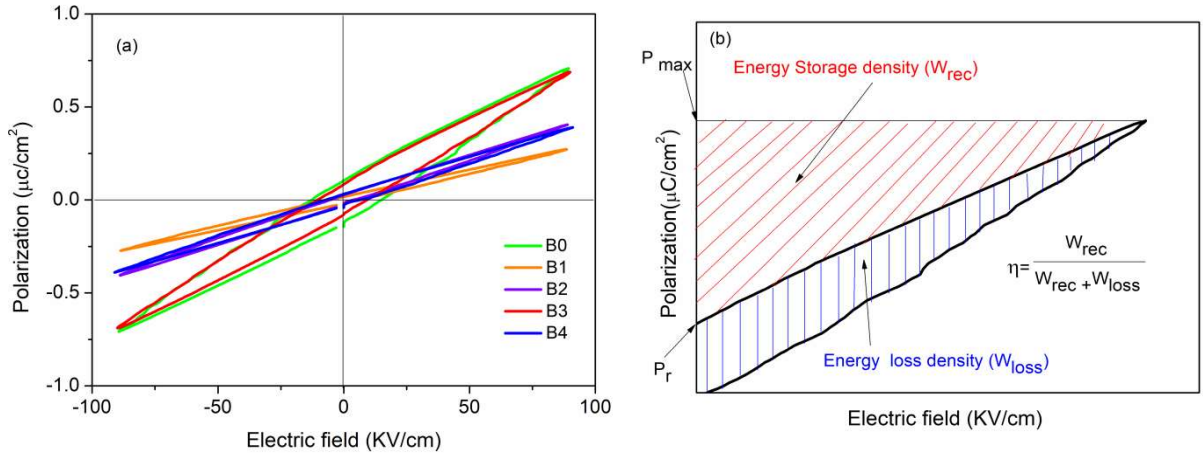

**Figure 9.** (a) *P–E* loops of all the investigated samples at electric field of 90 kV/cm, (b) schematic illustration of the calculation method of the energy storage properties from *P-E* loop.

**Table 3**. The evaluation of discharge energy density, the loss energy density and energy efficiency of all samples.

| Sample | $W_{rec}$(mJ/cm$^3$) | $W_{loss}$(mJ/cm$^3$) | $\eta$(%) |
|---|---|---|---|
| B0 | 25.8 | 11.9 | 68.35 |
| B1 | 11.2 | 4.4  | 84.83 |
| B2 | 16.6 | 2.6  | 86.39 |
| B3 | 25.6 | 9.1  | 73.77 |
| B4 | 16.0 | 3.4  | 82.30 |

It is noted that the *P-E* loops decreased with an increment concentration of $WO_3$, which is in agreement with the variation of dielectric loss at room temperature shown in figure 6. This decrease could be attributed to the increase of dielectric loss at room temperature. The recoverable energy ($W_{rec}$) for different substitution fraction x at electric field of E = 90 kV/cm are 25.8 mJ/cm$^3$, 11.2 mJ/cm$^3$, 16.6 mJ/cm$^3$, 25.6 mJ/cm$^3$, 16 mJ/cm$^3$ corresponding to B0, B1, B2, B3 and B4 respectively. The optimum recoverable density of 25.6 mJ/cm$^3$ was obtained for B3 with an energy efficiency of 73.77%.

The energy efficiency shows a behavior of increase with the rise in substitution fraction x. It reaches to a maximum value of 86.39% for B2.

It is worthy to notice that the values of discharge energy storage obtained in this work are comparable with the other silicate glass-ceramics reported in the literature [54, 55]. In addition, as it can be depicted form the plots, it seems that the *P-E* saturation could be achieved at very high values of applied fields that we have not been able to achieve with our experimental device.

## 4. Conclusion

The 2BaO–0.5Na$_2$O–1P$_2$O$_5$– 2.5[(1-x)Nb$_2$O$_5$- xWO$_3$] (x = 0–0.4) glass–ceramics samples were prepared via the melting method followed by controlling crystallization. The thermal analysis showed that the substitution of Nb$_2$O$_5$ by WO$_3$ depolymerizes the glass-ceramics. It was shown during this work that the nature of the formed phase influences the dielectric properties as well as the high temperature conductivity. This behavior was attributed to the existence of structural defect governed by the formed crystalline phase. The optimum recoverable density of 25.6 mJ/cm$^3$ was obtained for B3 with an energy efficiency of 73.77% under the electric field of 90kV/cm. Also, the energy efficiency was improved with the increase in substitution fraction. The ferroelectric properties show unsaturated hysteresis loops with low polarization at 90kV/cm and with energy storage properties comparable with other glass systems. We believe that our study could likely bring useful information to the researchers community deal with the energy storage properties in glass-ceramics, especially based on phosphate compound where the literature is still lacking.

 **Acknowledgments:**

The authors gratefully acknowledge the financial support of CNRST, OCP foundation and the European Union's Horizon 2020 research and innovation program ENGIMA under the Marie Skłodowska-Curie grant agreement No 778072.I.R.